\documentclass[article]{emulateapj}
\newcommand{\gal}{IRAS 04296}
\newcommand{\rth}{$\mathbb R_{13}$} 
\newcommand{\ret}{$\mathbb R_{18}$} 
\newcommand{\rthet}{$\mathbb R_{1318}$} 
 \usepackage{color}

\slugcomment{Accepted to the ApJ}
 
\shorttitle{Dense Gas in IRAS 04296+2923}
\shortauthors{Meier et al.} 
\begin{document} 
 
\title{Spatially Resolved Chemistry in Nearby Galaxies III. Dense Molecular Gas in the Inner Disk of the LIRG IRAS~04296+2923}

\author{David S. Meier\altaffilmark{1,2}, Jean L. Turner\altaffilmark{3}, and Sara C. Beck\altaffilmark{4}}

\altaffiltext{1}{Department of Physics, New Mexico Institute of Mining and 
Technology, 801 Leroy Place, Socorro, NM 87801, USA; dmeier@nmt.edu}
\altaffiltext{2}{National Radio Astronomy Observatory, P. O. Box O, Socorro, NM, 87801, USA}
\altaffiltext{3}{Department of Physics and Astronomy, UCLA, Los Angeles, 
CA 90095--1547, USA; turner@astro.ucla.edu}
\altaffiltext{4}{Department of Physics and Astronomy, Tel Aviv University,  
69978 Ramat Aviv, Israel; sara@wise.tau.ac.il}


\begin{abstract} 

We present a survey of 3 mm molecular lines in IRAS 04296+2923, one of the brightest known molecular-line emitting galaxies, and one of the closest LIRGs.  Data are from the Owens Valley and CARMA millimeter interferometers.  Species detected at $\lesssim 4\arcsec$ resolution include C$^{18}$O, HCN, HCO$^{+}$, HNC, CN, CH$_{3}$OH and, tentatively, HNCO.  Along with existing CO, $^{13}$CO and radio continuum data, these lines constrain the chemical properties of the inner disk. Dense molecular gas in the nucleus fuels a star formation rate $\gtrsim$10 M$_{\odot}$ yr$^{-1}$ and is traced by lines of HCN, HCO$^{+}$,  HNC, and CN.  A correlation between HCN and star formation rate is observed on sub-kpc scales, consistent with global relations.  Toward the nucleus, CN abundances are similar to those of HCN, indicating emission comes from a collection ($\sim$40-50) of moderate visual extinction, photon-dominated region clouds. The CO isotopic line ratios are unusual: CO(1--0)/$^{13}$CO(1--0) and CO(1--0)/C$^{18}$O(1--0) line ratios are large toward the starburst, as is commonly observed in LIRGs, but farther out in the disk these ratios are remarkably low ($\lesssim 3$). $^{13}$CO/C$^{18}$O abundance ratios are lower than in Galactic clouds, possibly because the C$^{18}$O is enriched by massive star ejecta from the starburst.  $^{13}$CO is underabundant relative to CO.  Extended emission from CH$_3$OH indicates that dynamical shocks pervade both the nucleus and the inner disk.   The unusual CO isotopologue ratios, the CO/HCN intensity ratio versus $L_{IR}$, the HCN/CN abundance ratio and the gas consumption time versus inflow rate, all indicate that the starburst in IRAS 04296+2923 is in an early stage of development.
\end{abstract} 
 
\section{Introduction \label{intro}} 

Luminous infrared galaxies (LIRGs) and ultraluminous infrared galaxies
(ULIRGs) represent the most active members of the star-forming galaxy
population in the nearby universe.  Often morphologically disturbed,
with vast reservoirs of atomic and molecular gas
\citep[e.g.,][]{SSEMMNS88,SSS91,SDRB97}, their high IR luminosities
are driven by prodigious amounts of efficient star formation
\citep[e.g.,][]{SM96}.  LIRGs are important laboratories for the
interplay of molecular gas and starbursts and their evolution over
time.  But locally LIRGs are rare \citep[e.g.,][]{SMNDELR87,SMKSS03},
so there are few examples that are near enough that we can detect and
resolve their chemical properties.
 
Despite being one of the nearest (29 Mpc; 140 pc =1\arcsec) LIRGs,
IRAS 04296+2923 [hereafter \gal]--- located behind the dark cloud
L1500 in Taurus --- has remained poorly studied until recently
\citep[][]{MTBGTV10}.  It was not so long ago that it was even
identified as a galaxy \citep[][]{SHDYFT92,CKSYT95}. Within 35 Mpc
only the galaxies NGC 1068, NGC 1365, NGC 2146, NGC 4418 and NGC 7552
are as IR luminous as IRAS 04296, at $\sim 10^{11}~L_{\odot}$
\citep[Table \ref{GalT};][]{SMKSS03}.  IRAS 04296 is remarkable for
the compactness of its starburst, with an estimated star formation
rate of $\sim$10~$\rm M_\odot \, yr^{-1}$ originating within the
central $2^{''}$ ($<280$ pc), and $\sim$25~$\rm M_\odot \, yr^{-1}$
for the entire galaxy \citep[][]{MTBGTV10}. CO mapping with the Owens
Valley Millimeter Array (OVRO) revealed an extremely massive molecular
gas disk of M(H$_{2}) \simeq 6\times 10^{9}~ \rm M_{\odot}$ within the
central 45$^{''}$ \citep[][]{MTBGTV10}.  No galaxy within 30 Mpc,
observed in the Five Colleges Radio Astronomy Observatory (FCRAO) CO
survey \citep[][]{FCRAO95}, has as high a CO luminosity over that
aperture.  Toward the central region $\rm H_2$ column densities are
$1\times 10^{23}~\rm cm^{-3}$ averaged over 500 pc scales.  \gal\ is
one of the most gas-rich systems in the nearby universe.

 \begin{deluxetable}{lcl}
\tablenum{1}
\tablewidth{0pt}
\tablecaption{IRAS 04296+2923 Basic Data}
\tablehead{\colhead{Characteristic}     & \colhead{Value}       &
\colhead{Ref.}}
\startdata
Dynamical Center\tablenotemark{a}    &  $04^{h} 32^{m} 48^{s}.65\pm 1^{''}$ &1   \\
(kinematic) [$J2000$]  & $+29^{o} 29' 57.^{''}45\pm 1^{''}$   &   \\
2$\mu$m peak (2MASS)     & $04^{h} 32^{m} 48^{s}.60\pm 0.^{''}3$  &1   \\
~[$J2000$]  &$+29^{o} 29' 57.''49\pm 0.^{''}3$ & \\
V$_{lsr}$\tablenotemark{a}   &2086 kms$^{-1}$   &1   \\
Adopted Distance  & 29 Mpc  &1   \\
Position Angle\tablenotemark{a} & 252$^{o}$ & 1\\
Inclination & 50$^{o}$ & 1 \\
M$_{H_{2}}$($<3.5\arcsec$)\tablenotemark{b}  & $4.3 \times 10^{8}~M_{\odot}$& 1 \\
$\Sigma_{H_{2}}$($<3.5\arcsec$)\tablenotemark{b} & 550 M$_{\odot}$ pc$^{-2}$  &1 \\
M$_{dyn}$ ($<3.5\arcsec$)\tablenotemark{a} & $1.6\times 10^{9}~M_{\odot}$ &  1 \\
M$_{H_{2}}$($<30\arcsec$)\tablenotemark{c}  & $5.9 \times 10^{9}~M_{\odot}$ & 1 \\
M$_{dyn}$ ($<30\arcsec$)\tablenotemark{a} & $3.0\times 10^{10}~M_{\odot}$ &  1 \\
$\rm M_{HI}^{tot}$ & $1.38 \times 10^{9}~M_{\odot}$& 3 \\
IRAS 12, 25, 60, 100$\mu$m& 1.39, 5.90, 42.1, 48.3 Jy &2\\
L$_{IR}$  & $9.8 \times 10^{10}~L_{\odot} $&  2 \\
\enddata
\tablenotetext{a}{Based on the best fitting rotation curve}
\tablenotetext{b}{From $^{13}$CO(1--0).}
\tablenotetext{c}{Assuming the standard CO conversion factor.}
\tablerefs{(1) Meier et al. (2010); (2) Sanders et al. (2003); (3) Chamaraux 
et al. (1995)} 
\label{GalT}
\end{deluxetable}

\begin{figure*}
\epsscale{1.0}
\plotone{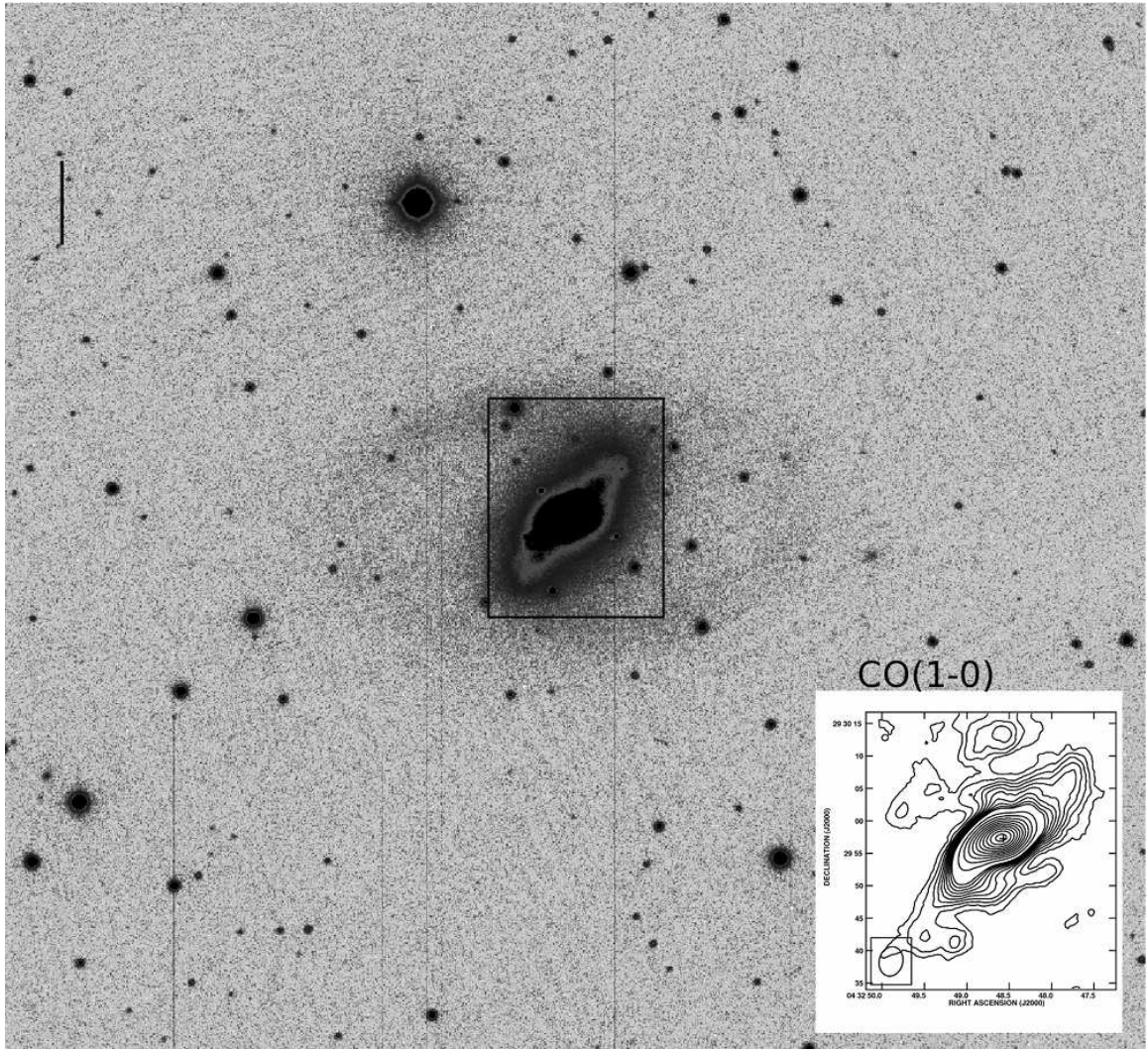}
\caption{The J band near infrared image of \gal\ from the Palomar 5m with CO(1--0) integrated 
intensity in the bottom right corner \citep[][]{MTBGTV10}.  The black box marks the field of 
view covered in CO(1--0).  Contours are  1, 2, 3, ... 10, 15, 20, ... 60 $\times$13.8 K km s$^{-1}$ for a 
beam size of $4.6\arcsec \times 3.6\arcsec ;-14^{o}$ \citep[][]{MTBGTV10}. The scale bar in the upper left is 2.8 kpc (20$\arcsec$) long.  
\label{Igal}}
\end{figure*}

In many respects, \gal\ appears to be ``normal'' bar-induced star
formation taken to an extreme \citep[Figure \ref{Igal};][]{MTBGTV10}.
Unlike many [U]/LIRGs, there are no obvious signs, morphological or
kinematic, of a recent major merger.  It is a symmetric, barred spiral
galaxy with a pronounced outer ``theta" ring.  However, judging from
the estimated gas inflow rates along the bar, \gal\ is not in a stable
configuration.  The nuclear star formation rate is so large that the
nuclear component is being consumed more rapidly than it can be
replenished by the bar-driven inflow of $\sim5~\rm M_\odot \,
yr^{-1}$. This argues that \gal\ is early in its LIRG state; the
observed starburst can be maintained for the next 100 Myr
\citep[][]{MTBGTV10}.

\gal 's unusual CO brightness raises questions about the dense gas in
this LIRG. How much dense gas is there in \gal, and where is it found?
What are the densities and temperatures of the gas clouds?  What are
the effects of the nuclear starburst on the dense gas?  What are the
effects of bar inflow on the dense gas?  What does the dense gas and
its chemistry reveal about the nuclear starburst in this LIRG?  Here
we use 3 mm aperture synthesis observations to address these questions
in \gal.  The target molecules are tracers of dense gas and probe a
range of chemical conditions.

\section{Observations \label{obs}} 

Aperture synthesis observations were obtained for \gal\ in dense gas
tracers at $\lambda$ = 3 mm with the Combined Array for Research in
Millimeter Astronomy (CARMA) \citep[][]{CARMA04}.  Table \ref{ObsT}
lists the molecular transitions surveyed along with their
observational parameters.  Observing parameters for data from the
Owens Valley Millimeter Array \citep[OVRO:][]{OVRO94} consisting of
transitions between 109-113 GHz are as reported in
\citet[][]{MTBGTV10}. The CO(1--0) and $^{13}$CO(1--0) observations
have a velocity resolution of 10.5 km s$^{-1}$. The lines presented
here were observed simultaneously in wideband mode (128$\times$31.25
MHz) giving a velocity resolution of $\sim$ 90 km s$^{-1}$.  The 88 -
97 GHz tuning was observed separately with the CARMA interferometer
but had the same wideband 31.25 MHz channels ($v_{chan}~\sim 100$ km
s$^{-1}$).  All detected lines are resolved spectrally (emission in at
least 3 adjacent channels) toward the nucleus, but the low spectral
resolution may imply some spectral `beam dilution' toward the narrower
line width disk.

The OVRO and CARMA data sets were calibrated using the MMA and MIRIAD
software packages, respectively.  Phase calibration was done by
observing the quasar J0336+323 every 25 minutes (OVRO) and 0237+288 or
0530+135 every 20 minutes (CARMA).  Absolute flux calibration was done
using Uranus as primary flux calibrator and 3C273, 3C84 and 3C454.3
for secondary flux calibration (both OVRO and CARMA).  Uncertainties
in absolute flux calibration are $\sim$10\% for both data sets.
Mapmaking was done in MIRIAD and subsequent data analysis and
manipulation was done with the NRAO AIPS package.  All data were
naturally weighted.  Spatial resolutions are $\lesssim 4.5^{''}$ for
the high frequency (OVRO) tuning and $\lesssim 3.3^{''}$ for the low
frequency (CARMA) tuning.  Integrated intensity images are moment 0
maps with all emission brighter than 1.3$\sigma$ per channel included.
Since the emission observed from these higher density tracers remain
confined well inside the half power point of the array(s), corrections
for the primary beam attenuation have not been applied.  No
single-dish observations of this galaxy exists for these transitions,
so no estimate of the amount of resolved-out flux is possible;
however, it is not expected that there is missing flux as it would
require the existence of a uniform dense component extended on
$\gtrsim$6 kpc scales.

\begin{deluxetable*}{lccccccc} 
\tabletypesize{\scriptsize}
\tablenum{2} 
\tablewidth{0pt} 
\tablecaption{Observational Data\label{ObsT}} 
\tablehead{ 
\colhead{Transition}  
&\colhead{Dates}
&\colhead{Frequency}
&\colhead{T$_{sys}$}
&\colhead{$\Delta V_{chan}$} 
&\colhead{Beam}
&\colhead{K/Jy}  
&\colhead{Noise} \\ 
\colhead{}  
&\colhead{\it (MMYY)}
&\colhead{\it (GHz)}  
&\colhead{\it (K)}
&\colhead{($km~s^{-1}$)} 
&\colhead{\it ($^{''}\times^{''}$;$^{o}$)}  
&\colhead{}
&\colhead{\it (mJy bm$^{-1}$)}}  
\startdata 
CARMA\tablenotemark{a}: & &  & & & & & \\
HCN(1--0) & 0808-1008&  88.632 & 130-230& 105.7 &$3.3\times 2.7$;-90 & 17.5 & 2.9 \\ 
HCO$^{+}$(1--0)& & 89.189 && 105.0&& 17.3&  2.9\\ 
HNC(1--0) & &  90.664 & & 103.3 & & 16.7 & 2.9 \\ 
C$^{34}$S(2--1)& &  96.413 & & 97.17 && 14.8 & 3.0 \\ 
CH$_{3}$OH($2_{k}$--$1_{k}$)& &  96.741 & & 96.84 & & 14.7 & 3.0 \\
OVRO\tablenotemark{b}: & & &  & & & & \\
HC$_{3}$N(12--11) & 1103-0504& 109.174 &220-450 & 85.8 &$4.7\times 3.8$;-15 & 5.91 & 4.0 \\ 
C$^{18}$O(1--0)\tablenotemark{c} & & 109.782 && 85.3 & & 5.86 & 4.0 \\ 
HNCO($5_{05}$--$4_{04}$)\tablenotemark{c}& &109.905  &  &  & & &  \\ 
CN(1--0;$\frac{3}{2}-\frac{1}{2}$)& &  113.491 & & 82.5 &$4.6\times 3.6$;-14& 5.71 & 4.5 \\ 
CN(1--0;$\frac{1}{2}-\frac{1}{2}$)\tablenotemark{d}& & 113.191& & & & &  \\ 
\enddata 
\tablenotetext{a}{Phase Center: $\alpha = 04^{h} 32^{m} 48^{s}.6~~ 
\delta = +29^{o} 29' 57.^{''}5$ (J2000)} 
\tablenotetext{b}{Phase Center: $\alpha = 04^{h} 32^{m} 48^{s}.6~~ 
\delta = +29^{o} 29' 58.^{''}0$ (J2000)} 
\tablenotetext{c}{Partially blended} 
\tablenotetext{d}{Observed simultaneously with CN(1--0;$\frac{3}{2}-\frac{1}{2}$)}
\end{deluxetable*} 

\section{Results: Integrated Intensity Maps and Abundances}  \label{res} 

\begin{figure*}
\epsscale{1.0}
\plotone{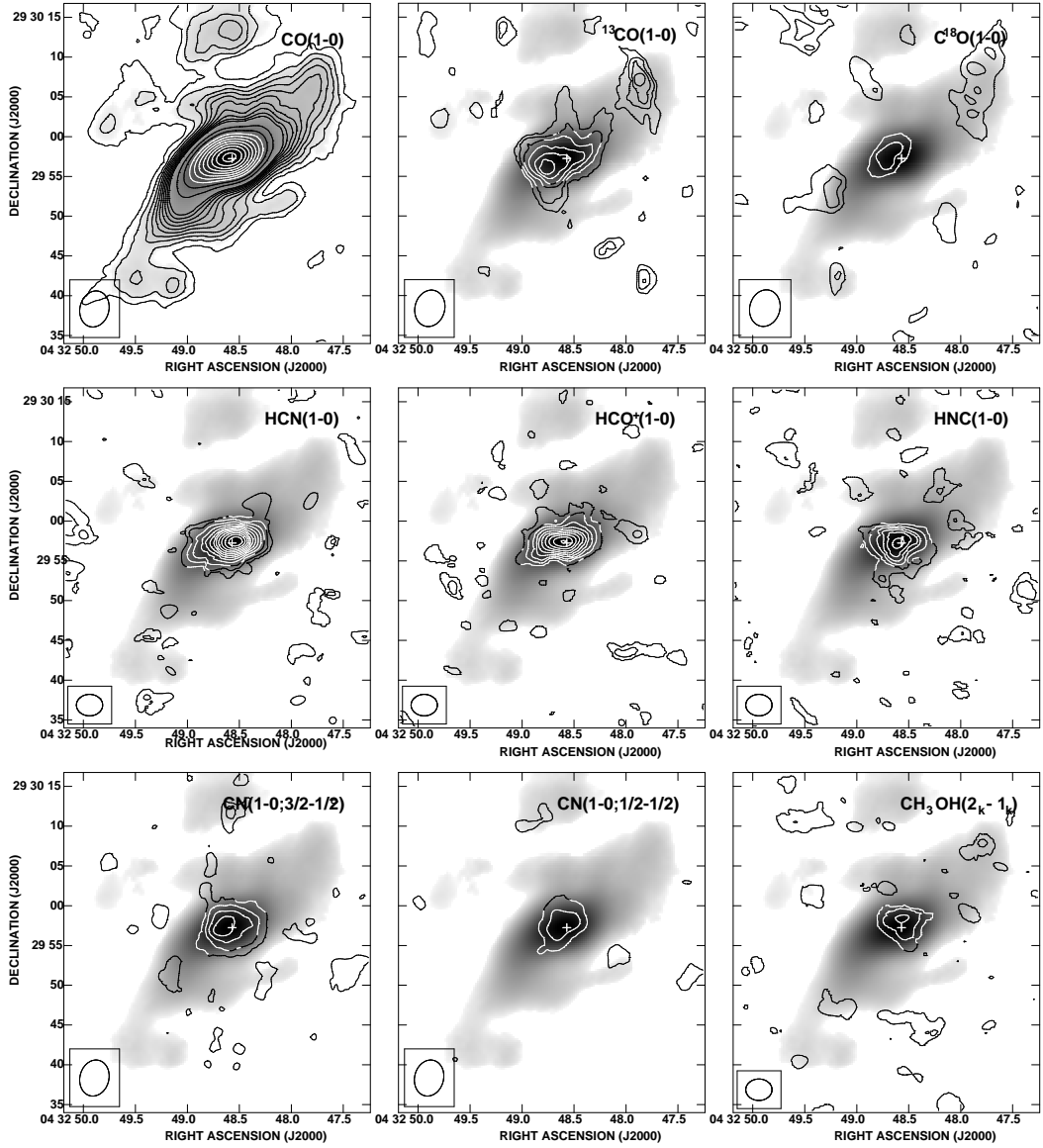}
\caption{Integrated intensity maps of the seven detected transitions, plus CO(1--0) in greyscale and 
$^{13}$CO(1--0).  The latter two are from  \citet[][]{MTBGTV10}.   The beam of all transitions is displayed in the lower left of each panel, and
 the cross marks the location of the starburst \cite[radio continuum peak;][]{MTBGTV10}. 
{\it Top Left)}   CO(1--0).  Contours are  1, 2, 3, ... 10, 15, 20, ... 60 $\times$13.8 K km s$^{-1}$ for a beam size of $4.6\arcsec \times 3.6\arcsec ;-14^{o}$.  
{\it Top Center)}  $^{13}$CO(1--0);  contour levels are 7.3 K km s$^{-1}$ for the same beam size as CO(1--0).   Beam sizes for the remaining transitions are as 
listed in Table \ref{ObsT}.
{\it Top Right)}  C$^{18}$O(1--0).    Contours  as for $^{13}$CO(1--0). 
{\it Middle Left)}  HCN(1--0). Contour levels are 13.1K km s$^{-1}$.  
{\it Middle Center)}  HCO$^{+}$(1--0). Contour levels are 17.3 K km s$^{-1}$.  
{\it Middle Right)}  HNC(1--0). Contours levels 16.7 K km s$^{-1}$. 
{\it Bottom Left)}   CN(1--0;$\frac{3}{2}-\frac{1}{2}$). Contours levels are 12.8 K km s$^{-1}$ 
{\it Bottom Center)}  CN(1--0;$\frac{1}{2}-\frac{1}{2}$).  Contours as 
for CN(1--0;$\frac{3}{2}-\frac{1}{2}$).  
{\it Bottom Right)}  Combined intensity for the 
CH$_{3}$OH($2_{k}-1_{k}$) quadruplet of lines. Contours levels are 16.2 K km s$^{-1}$.  The above contours 
correspond approximately to 2$\sigma$ increments for each transition except CO(1--0) (see Table \ref{GalT}).
\label{IntI}}
\end{figure*}

Figure \ref{IntI} displays the integrated intensity maps of the
detected transitions along with $^{12}$CO(1--0) and $^{13}$CO(1--0)
from \citet[][]{MTBGTV10}.  The CO(1--0) intensity map displays the
overall structure of \gal.  Two barred arms extend from the outer, low
pitch angle spiral arms into the central region where the gas collects
into a very bright nuclear feature, referred to as the `circum-nuclear
zone' (CNZ).  The CNZ has a radius, corrected for inclination,
$\simeq0.9\rm ~ kpc$ \citep[][]{MTBGTV10}.  The CNZ region is also the
site of an intense starburst which dominates the radio continuum and
mid-infrared emission \citep[][]{MTBGTV10}.  The compact core of the
starburst traced with 6 cm radio continuum (marked in Figure
\ref{IntI} by a cross) is confined to the inner 3\arcsec\ or $\sim$200
pc radius.  The centroid of this compact starburst is close to the
center of the CNZ dense gas distribution, but appears shifted slightly
($\sim1$\arcsec) southwest of the centroid of the CO.  Somewhat weaker
star formation traced by 20 cm radio continuum matches the extent of
the CNZ (see section \ref{densesfr}).  Beyond the CNZ, CO emission
extends out to a galactocentric radius of $\sim 25\arcsec$ or
$\sim$3.5 kpc.  CO(1--0) from the outermost portion of the field,
including the separate northernmost clump, originates in the outer
spiral arms.

Below we discuss the morphology of each dense gas transition. In the
following sections we adopt an excitation temperature, T$_{ex}$=30 K
for the nucleus and T$_{ex}$=10 K for the bar and arms.  These are
probably reasonable estimates given that the starburst is strongly
localized to the nucleus.  However, until multi-line studies can be
executed, these should be considered `reference values' only.

{\bf C$^{18}$O(1--0) and HNCO(5$_{05}$--4$_{04}$) ---} The
C$^{18}$O(1--0) and HNCO($5_{05}-4_{04}$) transitions appear together
in the same spectral window.  The transitions are separated by about
330 km s$^{-1}$, so may potentially be blended, but the velocity field
of the galaxy and the faintness of HNCO allows C$^{18}$O(1--0) to be
unambiguously separated.  C$^{18}$O(1--0) is surprisingly extended.
Like the (continuum subtracted) $^{13}$CO(1--0), C$^{18}$O(1--0) peaks
just southeast of the starburst. The CNZ is not significantly brighter
than the arms in C$^{18}$O(1--0), remarkably different from CO(1--0),
where it is more than an order of magnitude stronger.  This would be
consistent with higher cloud temperatures in the CNZ, which would
increase the intensity of optically thick CO.  Along the arms,
C$^{18}$O(1--0) brightnesses are comparable to $^{13}$CO(1--0).  At
the northern bar end $^{13}$CO(1--0) intensities increase slightly
compared to C$^{18}$O.  Peak brightnesses reach 0.10 K averaged over
$\sim$0.5 kpc scales near the starburst and close to this along the
northern arm.  C$^{18}$O column densities are discussed in detail
below (section \ref{isotope}).

HNCO($5_{05}-4_{04}$) is tentatively detected ($\sim 3\sigma$) only in
the western CNZ. For T$_{ex}$ = 30 K, the HNCO column densities peak
at N(HNCO) $\lesssim 2.3 \times 10^{14}$ cm$^{-2}$.  Increasing
T$_{ex}$ to 50 K would raise the N(HNCO) limit to $4.1 \times 10^{14}$
cm$^{-2}$ (Table \ref{MolT}).

{\bf HCN(1--0), HCO$^{+}$(1--0) and HNC(1--0) ---} All three of these
dense gas tracers are bright and compact.  HCN(1--0) is the brightest
with antenna temperatures peaking at 0.94 K, while HCO$^{+}$(1--0) and
HNC(1--0) are $\sim$ 15 \% and $\sim$ 50 \% fainter, respectively.
HCN(1--0) and HCO$^{+}$(1--0) peak toward the starburst but extend
beyond it to cover the entirety of the CNZ (Figure \ref{hcn20cm}).
The dense gas tracers have abundances ranging from $3.3 - 5.4\times
10^{-9}$ here, assuming optically thin emission with HNC and HCO$^{+}$
at the low end and HCN at the high end.  For HNC and HCO$^{+}$,
abundances are higher than usual for single-dish measurements toward
nearby star forming galaxies \citep[][]{HHMBWM95,NJHTM92}, but typical
of higher resolution interferometer values
\citep[][]{SFB98,KMVOSOIK01,MT05,KWWBRM07,MT12}.  Very weak emission
is detected in HCN at two locations along the arms but clear emission
is not detected in HCO$^{+}$ and HNC.  The CO/HCN contrast ratio
between nucleus (7-9) and arms (6-9) is similar (section
\ref{densefrac}).

{\bf CN(1--0; 3/2--1/2) and CN(1--0; 1/2--1/2) ---} CN(1--0; 3/2--1/2)
is dominated by the CNZ.  Faint emission is tentatively detected along
the southeast arm and the northernmost clump, but not in the
northwestern arm.  Both fine structure components of CN(1--0) are
clearly detected toward the CNZ.  Hyperfine structure is not resolved
due to the low spectral resolution of the data.  The
CN(3/2--1/2)/CN(1/2--1/2) intensity ratio is 1.8$\pm$0.5 matching the
optically thin theoretical value of 2, within uncertainties.  For a
T$_{ex}$ = 30 K, peak CN column densities are N(CN) $= 9.0 \times
10^{14}$ cm$^{-2}$.

{\bf CH$_{3}$OH(2$_{k}$--1$_{k}$) ---} This line is composed of four
transitions, ($2_{-12}-1_{-11}$) E, ($2_{02}-1_{01}$) A++,
($2_{02}-1_{01}$) E, and ($2_{11}-1_{10}$) E, which are blended in
these spectra.  We refer to the combined spectral feature as the
$2_{k}-1_{k}$ transition of methanol.  Unlike the dense gas tracers,
CH$_{3}$OH($2_{k}-1_{k}$) does not peak exactly at the starburst, but
slightly to the north.  Emission is tentatively detected from the bar
ends.  CH$_{3}$OH abundances are $X(CH_{3}OH) \simeq 2.3\times
10^{-8}$ toward the CNZ and within 50 \% of this value at the ends of
the bar.  These abundances are quite large, reaching values comparable
to the highest values seen on $\sim$50 pc scales toward strong bar
shocks in nearby spirals \citep[e.g.,][]{MT05,MT12}.

{\bf Non-detections ---} HC$_{3}$N(12--11) and C$^{34}$S(2--1) were
searched for but not clearly detected anywhere across the field.
There is tentative evidence for HC$_{3}$N(12--11) from the western CNZ
but we do not consider it a detection. Upper limits for
HC$_{3}$N(12--11)/HCN(1--0) and C$^{34}$S(2--1)/HCN(1--0) are
$\lesssim$0.086 and $<$0.064 (2$\sigma$), respectively.  Abundance
limits (for T$_{ex}$=30 K) are $\lesssim5.6\times 10^{-10}$ and
$<7.7\times 10^{-10}$ for HC$_{3}$N and C$^{34}$S, respectively.  For
C$^{34}$S(2--1) this abundance is only weakly constraining, implying
$X(CS) < 2\times 10^{-8}$ for a $^{32}$S/$^{34}$S isotopic ratio of
$\sim$24 \citep[e.g.,][]{CHWLC96}. Implied HC$_{3}$N abundance limits
are $\sim$ 3 -- 5 times lower than toward IC 342 \citep[][]{MTS11},
but similar to abundances observed for M 82 and Maffei 2
\citep[][]{AMMMB11,MT12}.

\section{Discussion \label{disc}}

\begin{deluxetable}{lcc} 
\tablenum{3} 
\tablewidth{0pt} 
\tablecaption{Molecular Abundances\tablenotemark{a} \label{MolT}} 
\tablehead{ 
\colhead{Species}  
&\colhead{CNZ/Starburst\tablenotemark{b}}
&\colhead{Northern Arm\tablenotemark{c}} 
}
\startdata 
N(H$_{2}$) & $1.0\times 10^{23}$\tablenotemark{d} & $3.8\times 10^{22}$\tablenotemark{e} \\
HCN & $5.6\times 10^{-9}$ & $5.3\times 10^{-10}$ \\
HNC & $3.3\times 10^{-9}$ & $<6.3\times 10^{-10}$ \\
CN & $9.0\times 10^{-9}$ & $<2.2\times 10^{-9}$ \\
HCO$^{+}$ & $3.4\times 10^{-9}$ & $<3.4\times 10^{-10}$ \\
C$^{34}$S & $<7.7\times 10^{-10}$ & $<1.1\times 10^{-9}$ \\
HC$_{3}$N & $\lesssim 5.6\times 10^{-10}$ & $<3.4\times 10^{-10}$\\
CH$_{3}$OH & $2.3\times 10^{-8}$ & $\sim 1.4\times 10^{-8}$ \\
HNCO & $\lesssim 2.3\times 10^{-9}$ & $<1.7\times 10^{-9}$ \\
\enddata 
\tablenotetext{a}{All upper limits are $2\sigma$.} 
\tablenotetext{b}{Assumes an excitation temperature of 30 K for all transitions and optically thin emission.} 
\tablenotetext{c}{Measured at $\alpha = 04^{h} 32^{m} 47^{s}.9$;  
$\delta = +29^{o} 30' 07^{''}$ (J2000) and assumes an excitation temperature of 10 K for all transitions.} 
\tablenotetext{d}{Based on the N(C$^{18}$O) value, a favored nuclear [CO/C$^{18}$O] 
isotopologue ratio of 200 (section \ref{isosb}) and CO/H$_{2}$ = $8.5\times 10^{-5}$ 
\citep[][]{FLW82}.} 
\tablenotetext{e}{As in $d$, except for the favored northern bar value of [CO/C$^{18}$O] 
isotopologue ratio of 225 (section \ref{isobar})} 
\end{deluxetable} 

\subsection{$^{13}$CO and C$^{18}$O as Probes of  CO Gas Opacity and Column \label{isotope}}

The CO isotopologues $^{13}$C$^{16}$O (``$^{13}$CO") and
$^{12}$C$^{18}$O (``C$^{18}$O") are valuable probes of opacity and
isotopic abundance when compared to the most abundant isotopologue,
$^{12}$C$^{16}$O (``CO").  Under the LTE approximation, the isotopic
ratios provide a direct constraint on gas opacity:
\begin{equation}
\mathbb R_{i} \simeq \frac{(1 - e^{-^{12}\tau})}{(1 - e^{-^{i}\tau})}
\end{equation}
where $^{12}\tau$ is the CO(1--0) optical depth and $^{i}\tau$ is the
CO(1--0) isotopologue optical depth ($^{13}\tau$ and $^{18}\tau$)
\citep[e.g.,][]{ABBJ95}. The lower opacity of the isotopologues also
allow us to study the bulk of the molecular gas that is not sampled by
optically thick CO.  Furthermore, comparisons of the isotopologues
with CO constrain gas opacity and isotopic abundance ratios and can
reveal non-LTE gas excitation \citep[e.g.,][]{MTH00,MT04}.  Typical
values for the \rth $\equiv$ CO(1--0)/$^{13}$CO(1--0) line intensity
ratio range from \rth\ $\sim$ 4--7 for Galactic disk clouds; this
range reflects both opacity effects and isotopic abundance ratios
varying from [CO/$^{13}$CO] $\sim$ 25--90 within the Galaxy.  The
inferred isotopic abundance ratio, [CO/$^{13}$CO], has its lowest
values in the Galactic center and increases with galacto-centric
radius, reaching $\sim$70 at the solar circle and $\gtrsim$120 in the
outer Galaxy \citep[e.g.,][]{WR94,MSBZW05, WB96}. A similar gradient
is observed in the Galaxy for [CO/C$^{18}$O], with [CO/C$^{18}$O]
$\sim$ 250 at the Galactic Center and $\sim$500 at the solar radius.
This implies [$^{13}$CO/C$^{18}$O] between 6 -- 10 across the Galaxy.
In external galaxies a wider range is seen, \rth\ $\sim$ 3 -- $>$30,
with the higher values tending to originate from LIRGs and ULIRGs
\citep[][]{AJBB91,CDC92,ABBJ95}.  Typical values for the \ret $\equiv$
CO(1--0)/C$^{18}$O(1--0) line intensity ratio are \ret\ $\sim$ 15-100.
$^{13}$CO(1--0)/C$^{18}$O(1--0) intensity ratios, denoted \rthet, tend
to be lower than the Galactic [$^{13}$CO/C$^{18}$O], having values of
3 -- 6 \citep[e.g.,][]{ABBJ95}.

\begin{figure*}
\epsscale{1.0}
\plotone{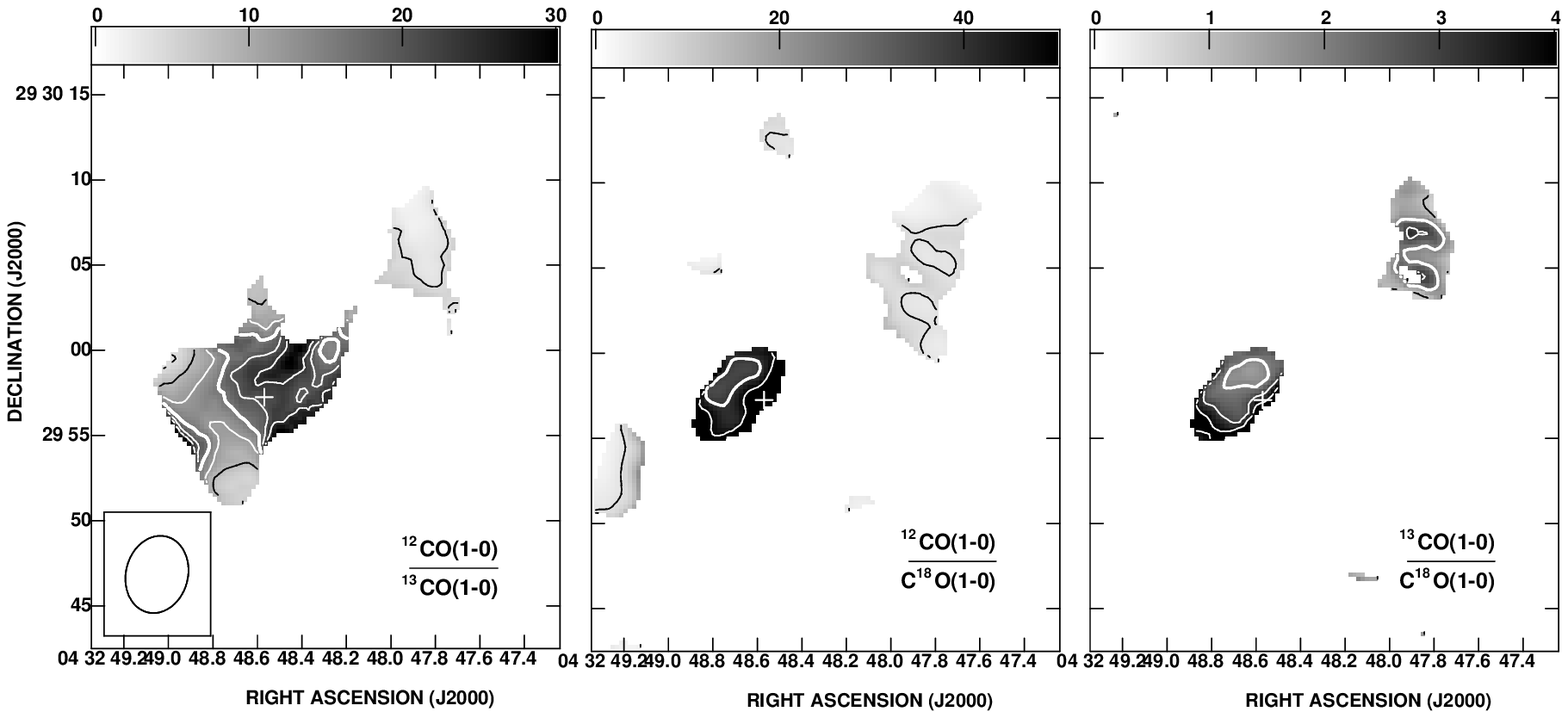}
\caption{The CO isotopologue ratios for \gal. {\it Left)} The CO(1--0)/$^{13}$CO(1--0)
line ratio \citep[][]{MTBGTV10}.   Contours are 4, 8, 12, 16 (bold contour), 20 and 24.
The grayscale ranges from 0 to 30 with dark being high ratios.  {\it Center)}  The
CO(1--0)/C$^{18}$O(1--0) intensity ratio with the CO(1--0) data from
\citep[][]{MTBGTV10}.  Contours are 6, 40 (bold contour), 50.  
The grayscale ranges from 0 to 50 with dark being high ratios.   {\it Right)} The
$^{13}$CO(1--0)/C$^{18}$O(1--0) intensity ratio with the $^{13}$CO(1--0) data from 
\citep[][]{MTBGTV10}.  Contours are 1, 2 (bold contour), 3, 4.  The 
grayscale ranges from 0 to 4 with dark being high ratios.  The beam size of all plots are 
the same and given in the lower left of the figure.
\label{Isorat} }
\end{figure*}

Figure \ref{Isorat} presents the three isotopic line ratios, \rth,
\ret, and \rthet.  All three exhibit the same trend that they are high
toward the CNZ and lower along the arms.  \rth\ is $16\pm 4$ toward
the CNZ but only $1.9\pm0.4$ at the north end of the northern arm (see
Table \ref{MolT} for position).  Along the inner part of the bar arms
\rth\ $\simeq$ 3-4.  Likewise \ret = $45\pm10$ toward the CNZ and
drops to $3.3\pm0.7$ by the end of the northern arm.  The double rare
isotopic ratio \rthet, is fairly low everywhere across the mapped
region, being $3.2\pm0.7$ toward the CNZ and falling to
\rthet\ $\simeq 1.7\pm0.5$ by the end of the northern bar.  toward the
starburst site, \rth\ (\ret) is 21$\pm$4 (94$\pm$30), even larger than
seen elsewhere in the CNZ.  However, \rthet\ is not significantly
altered at the starburst (\rthet = 3.7$\pm$1.3).

The elevated \rth\ and \ret\ ratios in the CNZ are not unexpected,
since this has long been seen in starburst regions
\citep[e.g.,][]{AJBB91,CDC92}. Common explanations for the elevated
\rth\ and \ret\ include, 1) lowered gas opacity due to broader line
widths that result in lower CO column densities per unit velocity, 2)
non-LTE effects that raise CO brightness relative to the
isotopologues, such as sub-thermal gas densities or PDR/externally
heated clouds, or 3) anomalous isotopic abundances.

In contrast, the low isotopic line ratios seen toward the bar arm
imply (for LTE) high opacities.  The isotopic ratios approach unity in
the limit of infinite opacity.  For Galactic local ISM abundance
ratios \citep[e.g.,][]{WR94}, $^{18}\tau \gtrsim 1$ would be required
to explain the very low \rthet\ values seen along the bar arms.
However, such an interpretation conflicts with the other two ratios.
If gas opacity is high enough to explain \rthet\ then both \rth\ and
\ret\ should exhibit ratios much closer to unity. The observed
\rth\ and \ret\ along the northwestern arm, while lower than toward
the nucleus, are still significantly above unity.

Possible explanations for these unusual ratios for the nucleus and the
bar arms are discussed below.

\begin{figure*}
\epsscale{1.0}
\plotone{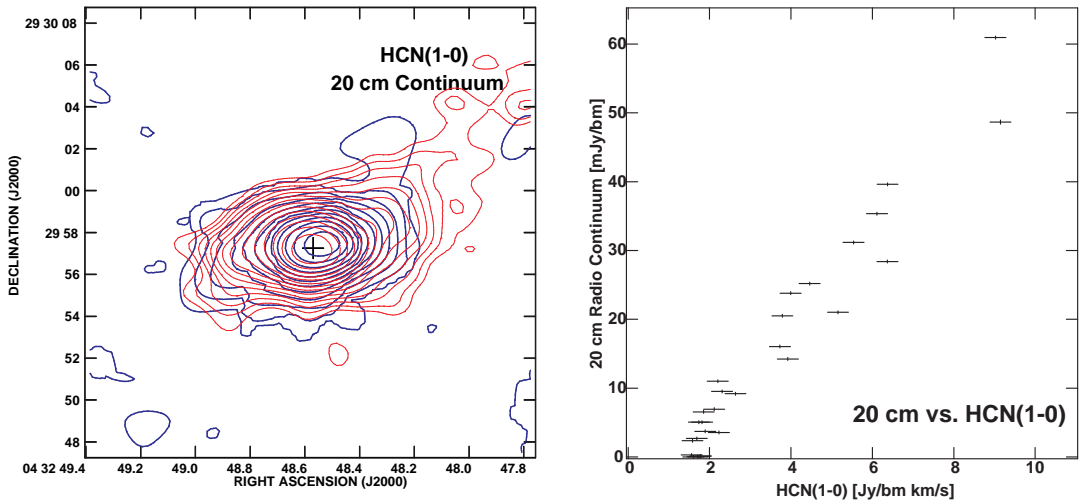}
\caption{{\it Left)} The HCN(1--0) integrated intensity maps overlaid on high resolution 20 cm radio 
continuum \citep[][]{MTBGTV10}.   Contours for HCN(1--0) are blue lines and are as in Fig. \ref{IntI}. 
Contours of 20 cm radio continuum are red lines in steps of 2$^{n/2}$, n=0,1,2 ... $\times$ 0.25 mJy 
beam$^{-1}$ for a resolution of $1.^{''}6 \times1.^{''}3$.  {\it Right)} The pixel by pixel correlation 
between HCN(1--0) and the 20 cm continuum.  The 20 cm image has been convolved to the same 
beamsize as the HCN(1--0) data and then sample in 1.5$^{''}$ ($\sim$ half beam) intervals.  Note that 
this sampling interval is smaller than the beam and so different plotted points are correlated.  Points with 
upper limits for both 20 cm and HCN(1--0) have been suppressed. 
\label{hcn20cm} }
\end{figure*}

\subsubsection{Low Opacity and Columns Toward the Circum-Nuclear Zone \label{isosb}}

The high ratios toward the CNZ are not extreme compared to other LIRGs
and imply low opacity CO gas.  A range of LTE solutions for gas
opacity and isotopic abundance are possible.  The optical depth of
CO(1--0), $^{12}\tau$, can range from $\simeq 4.0$, (corresponding to
[CO/$^{13}$CO] $\simeq 60$, and [CO/C$^{18}$O] $\simeq$200) to $\simeq
8.0$, (for [CO/$^{13}$CO] $\simeq 120$ and [CO/C$^{18}$O]
$\simeq$375).  Pushing $^{12}\tau$ larger than 8.0 requires very large
[CO/$^{13}$CO] abundance ratios, greater than 120.  These values would
be consistent with the low nuclear processing levels observed for gas
in the outermost parts of the Galaxy, based on the \citet[][]{WR94}
extrapolated gradient \citep[there are indications that this gradient
  may be too shallow in the far-outer Galaxy;][]{WB96}, and with
values of [$^{12}$C/$^{13}$C] $\sim 100$ recently suggested for the
centers of local starburst galaxies
\citep{MAMM10,2014A&A...565A...3H}.  On the other hand, pushing
$^{12}\tau$ lower than 4.0 requires [CO/C$^{18}$O] abundance ratios
lower than seen anywhere in the Galaxy and lower than the fully
massive star enriched values predicted from chemical evolution models
\citep[][]{HM93}.  Toward the starburst site, these constraints are
even more dramatic if LTE applies (see below).  Overall, low values of
$^{12}\tau$ are favored for the CNZ based on a comparison with the
northern arm (section \ref{isobar}).

Low CO opacities in CO-bright nuclear starburst may seem surprising.
Even $^{13}$CO has moderate, not low opacity in the more normal
spirals, IC 342 and Maffei 2 \citep[][]{WJ90,MTH00,MT01,MTH08}.  The
presence of the starburst and its location at the center of a barred
potential alters the situation somewhat: the CO line width of the
nuclear emission is about twice the line widths of the bar ends.  The
broader line accounts for some of the increased CO intensity toward
the nucleus, but not all.  The C$^{18}$O peak brightness temperatures
toward the CNZ and the northern arm differ by less than a factor of
two.

Non-LTE effects can also be responsible for the high \rth\ and
\ret\ values, especially toward the starburst. The CO(1--0)/HCN(1--0)
and HCN(1--0)/HCO$^{+}$(1--0) line ratios over the CNZ argue that the
gas is dense enough that sub-thermal CO emission can be neglected
across the nucleus (section \ref{dense}).  However, the chemical data
indicate that the nuclear molecular emission is may be partially
influenced by photon-dominated regions (PDRs) (section \ref{chem}).
The radiation field from the starburst can heat the surfaces of the
nearby clouds, preferentially exciting the optically thick ($^{12}$CO)
transitions relative to the optically thin $^{13}$CO and C$^{18}$O.
The high radiation fields can also preferentially photo-dissociate the
optically thinner species.  Both mechanisms raise \rth\ and
\ret\ relative to their values in quiescent conditions
\citep[e.g.,][]{MTH00}.  Since $^{13}$CO is not highly opaque, there
should be a much weaker influence on \rthet.  Non-LTE effects are able
to explain the elevated ratios toward the starburst relative to the
CNZ, but must be extreme to change the conclusion that the CNZ has
modest $^{12}\tau$.

So even accounting for non-LTE effects associated with the starburst,
it appears that the observed nuclear CO isotopic line ratios imply
quite low $^{13}$CO opacities $^{13}\tau ~ \ll 1$, [CO/$^{13}$CO]
$\geq$60 and [CO/C$^{18}$O] $\gtrsim$ 200.  The isotopic abundances
ratios are consistent with an ISM enriched in $^{18}$O from recent
massive star ejecta.  Moreover, a relatively lower abundance of
$^{13}$C suggests less long term nuclear processing, since C is a
primary and $^{13}$C is a (mostly) secondary nucleus \citep[][]{HM93}.

Adopting [CO/C$^{18}$O] $\simeq$ 200 and an excitation temperature of
30 K along with a CO/H$_{2}$ abundance ratio of $8.5\times 10^{-5}$
\citep[][]{FLW82}, we derive an H$_2$ column density of N(H$_{2}$)
$\simeq 1.0\times 10^{23}$ cm$^{-2}$ toward the CNZ. This is about a
factor of two lower than predicted based on a Galactic conversion
factor of 2.0$\times 10^{20}$ cm$^{-2}$ (K km s$^{-1}$)$^{-1}$
\citep[][]{Het97,Set88}.  The Galactic conversion factor can be
accommodated either by adopting a nuclear excitation temperature of
$\sim$50 K over $\sim$200 pc scales, or raising [CO/C$^{18}$O] to
$\sim$340.

\subsubsection{Anomalous $^{13}$CO and C$^{18}$O Ratios Across the Bar Arms \label{isobar}} 

toward the bar arms the isotopic line ratios are very low compared to
what is typically observed for disks in other extragalactic systems
\citep[e.g.,][]{Pet01}.  Under LTE, low \rth\ and \ret\ imply large
$^{12}\tau$ ($\sim$65 -- 110).  It is counter-intuitive that the
northern arm would have gas opacity at least an order of magnitude
larger than the CNZ, while its CO intensity is more than an order of
magnitude fainter.  For these high opacities, the observed
\rthet\ imply [$^{13}$CO/C$^{18}$O] $\sim$2.1, lower than favored for
the CNZ and more than three times lower than found in either the local
Galactic ISM or the Galactic center region.

Assuming LTE it is difficult to obtain a consistent solution for these
line ratios.  To simultaneously match the three ratios at the end of
the northern arm we require [CO/$^{13}$CO] $\simeq$90 -- 140
(unusually high) and [CO/C$^{18}$O] $\simeq$ 185 -- 300, as
$^{12}\tau$ ranges from 65 -- 110.  Given the strong bar here it is
possible that radial inflow of relatively unprocessed outer disk gas
could explain the high [CO/$^{13}$CO] ratio observed for the bar arms,
however this explanation would imply that the [CO/C$^{18}$O] abundance
ratio should also be raised, which is not observed.  These
[CO/$^{13}$CO] ratios suggest that stellar processing on the longer
timescales typical of intermediate mass star lifetimes is rather low.

If we accept the high end of the opacity range for the CNZ and the low
end of the range for the end of the northern arm (to minimize dramatic
opacity differences) then the implied abundance ratios [CO/$^{13}$CO]
and [CO/C$^{18}$O] would both decline with galacto-centric radius.
This is at odds with the measured gradients in the Galaxy and chemical
evolution models.  To have [CO/$^{13}$CO] and [CO/C$^{18}$O] gradients
in the right sense we are forced to favor low opacity for the CNZ and
high opacity for the outer arms.  But even this is not particularly
satisfying because with high $^{12}\tau$ in the arms, [CO/C$^{18}$O]
$\simeq$ 300.  Moreover, if we adopt T$_{ex} \simeq$ 10 K and this
abundance the implied N(H$_{2}$) column is almost five times that
obtained using the Galactic conversion factor.  Such high conversion
factors may be seen in low metallicity systems, but not in solar
metallicity gas \citep[e.g.,][]{BWL13}

The anomalies in the ratios demand that we consider non-LTE effects.
The PDRs that were discussed in the nuclear region are not relevant in
the bar arms because strong star formation is absent.  Two other
non-LTE effects worth considering are, 1) sub-thermal excitation of
the isotopologues and 2) $^{13}$CO chemical fractionation in the arm
medium versus the CNZ.  The second we dismiss because we observe
[$^{13}$CO/C$^{18}$O] to be smaller than toward the CNZ.  Since
$^{18}$O is not expected to fractionate efficiently in the cold ISM,
fractionation would act to raise the [$^{13}$CO/C$^{18}$O] abundance
ratio relative to the nucleus, not lower it.  Moreover, detailed
studies find little evidence for $^{13}$CO fractionation actually
being observed in the ISM \citep[e.g.,][]{MSBZW05}.

Sub-thermal excitation is a viable explanation, especially since the
clouds in the arms likely have lower densities than the nuclear disk.
The large opacity of CO lowers its effective critical density relative
to the CO isotopologues by the escape probability factor ($\beta ~
\sim ~1/^{12}\tau$).  Therefore when gas densities drop below $\sim$
n$_{H_{2}} \leq ~10^{3.5}$ cm$^{-3}$, the brightness temperatures of
the CO isotopologues can drop relative to CO, inflating \rth\ and
\ret\ \citep[][]{MT01}.  This agrees with the observed behavior only
if the true \rth\ and \ret\ are unity ($^{12}\tau \rightarrow
\infty$).

If we adopt the $^{12}\tau \simeq$ 80 solution for the end of the
northern arm, the lowest that keeps the correct sense of the abundance
gradient, then [CO/$^{13}$CO] $\simeq$ 110 and [CO/C$^{18}$O] $\simeq$
225.  For an excitation temperature of 10 K, the observed $^{13}$CO
and C$^{18}$O intensities imply areal filling factors of $f_{a} \sim
0.03$.  This is reasonable for the bright, extended molecular gas disk
characteristic of \gal.  Adopting the above abundances and temperature
for the arm, N(H$_{2}) \simeq 3.8\times 10^{22}$ cm$^{-2}$ at the
outer end of the northern arm.  This column is larger than that from a
Galactic conversion factor but is not extreme; we will use it to
estimate abundances toward the northern arm. Until J = 2 -- 1
transitions are observed so the gas excitation can be directly
measured, implied conversion factors must be regarded as suspect, but
it is clear that the conversion factor for the CNZ is lower than in
the arms \citep[][]{MTBGTV10}, as it is in many galactic centers.

\subsubsection{CO Isotopologues and Stellar Processing in \gal}

We have seen that CO isotopic abundance ratios in \gal\ are anomalous
compared to the Galaxy and other local galaxies.  The observations
further show that the physical conditions of the molecular gas in the
arms are very different from that observed in the CNZ.  Taking into
account gas columns, absolute abundance ratios and their variation
with galacto-centric radius, we conclude that $^{12}\tau~\sim$ 4 -- 6,
[CO/$^{13}$CO] $\simeq$ 60 and [CO/C$^{18}$O] $\simeq$ 200 toward the
CNZ.  At the outer end of the bar arm these values have changed to
$\sim$80, 110, and 225, respectively.  The [CO/C$^{18}$O] ratios are
fairly low, consistent with the inner Galaxy value and with enrichment
from massive star ejecta over much of the inner $r ~\sim$3.5 kpc.
However, the inner disk [CO/$^{13}$CO] values are like those
$\gtrsim$10 kpc out in the Galactic disk.  Recent observations suggest
high $^{12}$C to $^{13}$C isotopic abundance ratios of $\rm
[^{12}C/^{13}C]\sim 100$ may be common in other starburst nuclei
\citep[][]{MAMM10,2014A&A...565A...3H}.  Hence if [CO/C$^{13}$O] can
be a proxy of longer term nucleosynthetic processing, the underlying
disk of \gal\ and possibly other starbursts are less processed than
most of the Galaxy.

\citet[][]{MTBGTV10} concluded from the molecular gas and dynamical
masses that \gal\ must be experiencing one of its first major
starburst episodes.  The isotopic abundances agree with this initial
burst scenario, but also show that the current burst is mature enough
to have enriched the nuclear disk and potentially the inner bar with
massive star ejecta. In this picture \gal's evolutionary state
parallels the Large Magellanic Cloud: a relatively unprocessed galaxy
with a current burst of massive star formation.  But the LMC appears
to have a completely different abundance pattern, being somewhat
enriched in $^{13}$CO, extremely depressed in C$^{18}$O
\citep[e.g.,][]{WCHWC09}, and with no sign of local isotopologue
variation across the galaxy (Hughes et al. 2014, in prep.).  Finding
the metallicity of \gal\ could help determine the stage of enrichment.

\subsection{The Dense Molecular Gas Component in \gal} 

CO(1--0) is a good tracer for overall molecular gas morphology, but it
is the dense gas from which stars form.  We need to observe molecular
species with higher critical densities than CO(1--0) to find the state
and characteristics of the star forming component of molecular gas.
HCN(1--0) is an example of a dense gas probe that linearly correlates
with star formation rate (SFR) \citep[][]{GS04}.  Other dense gas
probes include HCO$^{+}$, HNC, CS, HC$_{3}$N and CN.  Here we
investigate the nature of the dense gas as traced by these species and
their connection with the SFR.

\subsubsection{CO(1--0)/HCN(1--0) and the Dense Gas Fraction \label{densefrac}}

Since CO(1--0) traces low density molecular gas and HCN(1--0) high
density, the CO(1--0)/HCN(1--0) line ratio is a measure of the
fraction of gas that is dense.  The tight correlation observed between
HCN(1--0) and the SFR (section \ref{densesfr}), together with the
compactness of nuclear star formation in \gal\ \citep[][]{MTBGTV10},
suggests that the dense gas fraction ought to increase toward the
starburst.

This is indeed seen across the CNZ, with CO(1-0)/HCN(1--0) decreasing
from $\ge$12 at the outer edge of the nuclear disk to $\sim$7 at the
starburst.  Just southwest of the starburst the ratio even drops below
six, at the low end of that found by (single-dish) surveys of nearby
[U]/LIRGs \citep[e.g.,][]{SDR92,ABBJ95,GS04,BHLBW08}.  Even averaged
over the inner $20^{''}$ diameter, (close to single-dish sampling
scales), the CO(1--0)/HCN(1--0) line ratio is $<8$.  We see that the
fraction of dense molecular gas, $n_{H_2}\gtrsim 10^{4.5}$ cm$^{-3}$,
in the CNZ is high near the nuclear starburst, and falls away from the
starburst.

Along the arms HCN(1--0) is only tentatively detected in a few
locations (see Fig. \ref{denserat}), so there are just a few isolated
regions with elevated dense gas fractions.  Given the much fainter
CO(1--0) at these arm locations, they have CO(1--0)/HCN(1--0) limits
that are comparable to or slightly higher than the starburst values.
But the CO isotopologues are likely better tracers of gas column
(section \ref{isotope}). If $^{13}$CO/HCN or C$^{18}$O/HCN is used to
constrain the dense gas fraction then there is a pronounced decrease
in the dense gas fraction between the CNZ and the bar arms, as
expected given the lower star formation rate there.

\subsubsection{Dense Gas Properties of the Circum-Nuclear Zone \label{dense}}

Line ratios between HCN(1--0), HCO$^{+}$(1--0), HNC(1--0) and CN(1--0;
3/2--1/2) depend on gas physical conditions as well as chemistry.  The
dominant physical and chemical processes controlling the line
intensities of these transitions have been extensively discussed in
Paper II \citep[][]{MT12} and the literature \citep[e.g.,][]{APHC02,
  GGPC06, MSI07, P07, BHLBW08, KNGMCGE08, LSBM08, KMPIS12}, so here we
only briefly summarize.

The first-order physical parameter controlling these ratios is gas
density.  The critical density of HCO$^{+}$(1--0) is nearly and order
of magnitude lower than HCN(1--0), while HNC(1--0) has a critical
density slightly lower than HCN(1--0).  Therefore in the density range
from $10^{4-6}$ cm$^{-3}$, the HCN/HCO$^{+}$ depend on gas density,
with HCO$^{+}$ favored at lower densities relative to HCN.
Furthermore, HCO$^{+}$, being a molecular ion, has its abundance
decreased in high density gas due to faster recombination with
electrons \citep[e.g.,][]{P07}.  Together these two effects suggest
that HCN should be significantly brighter than HCO$^{+}$ in high
density gas.  In normal gas phase chemistry the HCN/HNC intensity
ratio is driven to unity through the mutual formation reaction
HCNH$^{+}$ + e$^{-}$ $\rightarrow$ HNC/HCN + H \citep[e.g.,][]{TH98,SHNI98}.
In environments that are hot or have experienced strong shocks or PDR
irradiation, HNC can be rapidly converted to HCN so this ratio can
deviate significantly from unity \citep[e.g.,][]{SWPRFG92,TPM97}.  Thus we
expect HCN/HNC intensity ratios to be near unity over a wide range of
high density gas conditions.  Where deviating from unity we expect HCN
to be strongly favored in hot, disturbed gas. This is generally
consistent with what we observe toward the more moderate starbursts,
IC 342 and Maffei 2 in Papers I and II \citep[][]{MT05,MT12}.  CN is
expected to trace PDR gas and its chemistry is discussed in more
detail in section \ref{pdr}.  Finally it should be noted that high
optical depth in these lines will act to hide any differences in the
physical and chemical behavior, driving all the ratios to unity.

\begin{figure*}
\epsscale{1.0}
\plotone{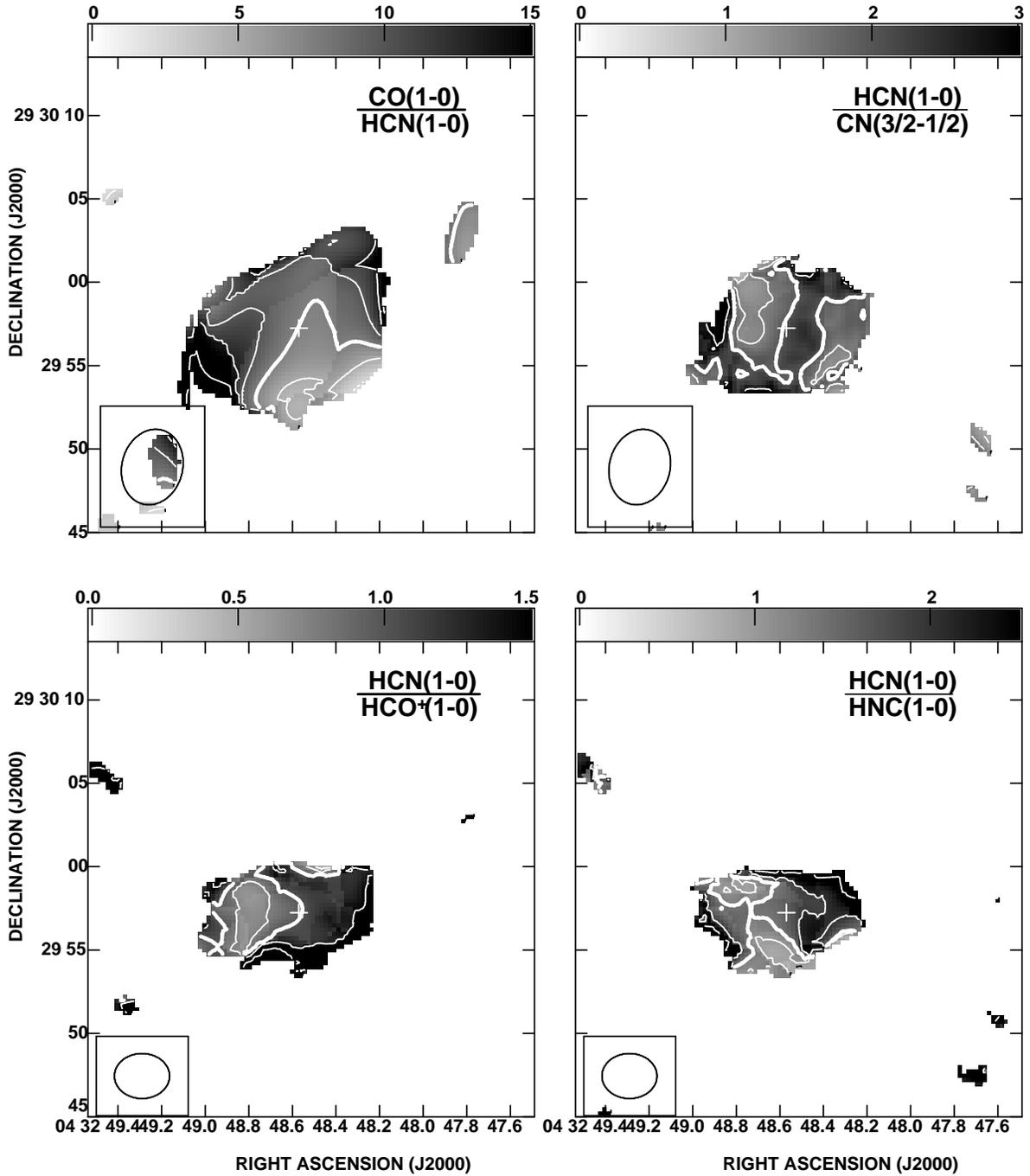}
\caption{The intensity ratios for dense gas tracers.  {\it Top Left)} CO(1--0)/HCN(1--0) line ratio.  
Contours are 5, 7.5 (bold contour), 10, 13.3, and 20.  The grayscale ranges from 0 to 15. 
In all planes darker grayscales correspond to higher ratios.    {\it Top Right)} The 
HCN(1--0)/CN(1--0;3/2-1/2) line ratio.  Contours are 1.25, 1.5, 2.0 (bold contour), and 3.0, with 
grayscale ranging from 0 to 3.  {\it Bottom Left)} HCN(1--0)/HCO$^{+}$(1--0) line ratio.  
Contours are 0.563, 0.75, 1.0 (bold contour), 1.5 and 3.0.  The grayscale ranges from 
0 to 1.5.   {\it Bottom Right)}  HCN(1--0)/HNC(1--0) line ratio.  
Contours are 0.75, 1.0, 1.33 (bold contour), 2.0 and 4.0.  The grayscale ranges from 0 to 2.5.  
In all planes the cross marks the location of the starburst (peak of the cm radio continuum 
\cite[][]{MTBGTV10}.  The beam for the ratios is displayed in the bottom left.  \label{denserat}}
\end{figure*}

Ratios of these three transitions are displayed in
Fig. \ref{denserat}.  Both the HCN(1--0)/HCO$^{+}$(1--0) and the
HCN(1--0)/HNC(1--0) line ratios exhibit the same east - west gradient
across the CNZ, with high values at the western side.
HCN(1--0)/HCO$^{+}$(1--0) ranges from 0.7 -- 1.5 and
HCN(1--0)/HNC(1--0) from 1.3 -- 2.5. The nuclear starburst resides
approximately in the middle of the observed ratio gradient in both
cases, so no clear evidence is seen for a distinct component directly
associated with the compact starburst.  The fact that the
HCN/HCO$^{+}$ intensity ratio is 1.0 toward the starburst suggests
that densities are high enough to thermalize both transitions and that
the opacities of both lines could be high. Large velocity gradient
modeling of HCN \citep[see][for model]{MTH08} and HCO$^{+}$ toward the
starburst (not shown) imply that for a kinetic temperature of 40 K
(the dust temperature, Table \ref{GalT}) densities are $\sim 2 \times
10^{5}$ cm$^{-3}$. The inferred HCN abundance (Table \ref{MolT}) is
near that typically observed on large scales toward star forming
clouds \citep[e.g.,][]{BSMP87}, so HCN opacities, while probably $>$1,
are unlikely to be extreme.

Both HCN/HCO$^{+}$ and HCN/HNC have the same sense of trend, rising
toward the west across the CNZ.  One possible model for these ratios
is that the gas density is lower in the east away from the starburst,
resulting in lower HCN/HCO$^{+}$ there, and warm PDR or shocked gas is
more prevalent in the west near the starburst, favoring HCN emission
and producing the high HCN/HNC line ratio there.  The starburst and CO
column density peak, is located at the inner terminus of the
northwestern bar arm, suggesting the western side of the nucleus to be
hotter / more energetic \citep[][]{MTBGTV10}.  In this model the
elevated HCN/HCO$^{+}$ is a better indicator of high gas density and
HCN/HNC more reflective of high radiation fields.  However opacity
must also be considered.  A second possible model is that line ratios
near the starburst could instead be dominated by high HCN line
opacity; if HCN is opaque but the two lower abundance species,
HCO$^{+}$ and HNC, transition from optically thick to thin from west
to east, following the decline in CO column, then the decline in
HCO$^{+}$ and HNC brightness relative to HCN toward the east could be
explained by their decline in opacity in this region.  However this
second model is less favored since HCN/HCO$^{+}$ (and HCN/HNC) on the
western side are significantly above one, ratios naively inconsistent
with high opacities.

\subsubsection{HCN(1--0) and the Nuclear Star Formation Rate \label{densesfr}}

Here we consider \gal, a spatially resolved LIRG, in the context of
the global $\rm L_{IR}$ versus HCN(1--0) relations
\citep[e.g.,][]{GS04,GUAGPCPA12}. The total HCN(1--0) luminosity over
the CNZ is L$_{HCN}$ = 7.4$\times10^{7}$ K km s$^{-1}$ pc$^{2}$.  This
equates to L$_{IR}$/L$_{HCN}$ = 1300 L$_{\odot}$ (K km s$^{-1}$
pc$^{2}$).  This value is 1.5 times the global values of [U]LIRGs
\citep[][with L$_{IR}\simeq 10^{11}$ L$_{\odot}$]{GS04, GUAGPCPA12}.
However the IR fluxes from IRAS cover a much larger area than the HCN
emission.  If we adopt L$_{IR}$(CNZ) $\simeq$ 0.5 L$_{IR}$(tot)
\citep[][]{MTBGTV10}, then the observed nuclear value for \gal\ is
L$_{IR}$/L$_{HCN}$=660, which is 40\% smaller than the global values.
That the CNZ in \gal\ is slightly IR under-luminous normalized by the
HCN compared to global averages is consistent with the determination
focused tightly on the dense gas of the CNZ region.

Since HCN(1--0) is imaged at high resolution, a rough estimate is
attempted to see if the correlation also persists within the galaxy.
\citet[][]{MTBGTV10} provide a detailed discussion of the rate,
efficiency and distribution of star formation over the nucleus of
\gal.  However, the high frequency radio data does not sample the SFR
on spatial scales comparable to HCN(1--0) and so we use a cruder SFR
proxy, the 20 cm radio continuum map.  Figure~\ref{hcn20cm} displays
the 20 cm radio continuum \citep[][]{MTBGTV10} compared with HCN(1--0)
at matched spatial resolution.  There is a clear linear correlation
between the two. Adopting a $q$ parameter value of 2.3 ($q$ = log$[\rm
  L_{FIR}/3.75 \times 10^{12} W/m^{2}] - \rm log[S_{20cm}/10^{26}
  Jy]$), typical of what is commonly observed for star forming
galaxies \citep[][]{C92}, the observed L$_{IR}$/L$_{FIR}$ ratio
\citep[with L$_{IR}$ being the total luminosity from 8 - 1000$\mu m$
  and L$_{FIR}$ being the luminosity from 40 - 400$\mu
  m$;][]{SMKSS03}, and \citet[][]{GS04}'s conversion between L$_{IR}$
and SFR allows the 20 cm map to be roughly converted to a resolved SFR
map.  The observed correlation between HCN(1-0) and 20 cm radio
continuum intensity is:
\begin{eqnarray}
\rm I_{20}(mJy ~bm^{-1}) &= [0.38\pm0.08] I_{HCN}(K ~km ~s^{-1})  \nonumber \\
 &~~~~~~~~- [0.42\pm0.35]
\end{eqnarray}
For the above normalization, after converting intensities to fluxes and luminosities, we find: 
\begin{eqnarray}
\rm SFR(M_{\odot} ~yr^{-1}) &= [1.4\pm0.3 \times 10^{-7}]~L_{HCN}(\rm K ~km s^{-1}~ pc^{2}) \nonumber \\
&~~~~~~- [0.02\pm0.02].  
\end{eqnarray}
The quoted errors are statistical only and do not include (potentially
large) systematic errors associated with uncertainties in adopted q
and the conversion between L$_{IR}$ and SFR.  Discounting the very
small zero-point offset, we find a slightly lower normalization than
the SFR(M$_{\odot}$ yr$^{-1}$) = ($1.8\times 10^{-7}$) L$_{HCN}$(K km
s$^{-1}$ pc$^{2}$) value of \citet[][]{GS04}.  Given the assumptions
this is considered good agreement between the local and global values.

HCO$^{+}$, which has a critical density lower than HCN's, shows a
statistically identical relationship.  \citet[][]{KT07} suggest that
the slope and normalization of the SFR vs. molecular gas tracer
relationship should differ between species that are fully thermalized
and those that are not.  The fact that we do not see a difference is
further evidence that the molecular gas localized to the starburst in
\gal\ has a density high enough to excite both HCN(1--0) and
HCO$^{+}$(1--0) equally well over much of the CNZ and that both may
have moderate opacity.  This finding is consistent with other
observations of dense gas tracers in nearby, resolved star-forming
galaxies \citep{2014ApJ...784L..31Z}.

\subsubsection{PDRs and the Nuclear Starburst in \gal\ \label{pdr}}

The interaction of UV photons from the starburst's massive stars with
the molecular gas is expected to result in PDRs with distinctive
chemical properties \citep[e.g.,][]{TH85, SD95}.  Given the strong
nuclear star formation in \gal, does it show signs of radiative
feedback from the starburst in the gas chemistry?

CN is expected to be abundant in PDRs.  It forms from reactions
between CH and N, with CH forming directly from H$_{2}$ and
PDR-abundant C$^{+}$\citep[]{SD95}.  CN(1--0; 3/2--1/2) and HCN(1--0)
have similar critical densities, so the HCN(1--0)/CN(1--0;3/2-1/2)
(hereafter HCN(1--0)/CN(1--0)) line ratio is considered to be an
excellent tracer of dense PDRs \citep[e.g.,][]{SD95,BS05}.

Toward the starburst in \gal\ the HCN(1--0)/CN(1--0) line ratio
reaches 2.4, which for LTE corresponds to an [HCN/CN] abundance ratio
of $\ge 0.6$ and a CN abundance of 9.0$\times10^{-9}$. (The abundance
ratio is a lower limit because of the potentially large HCN opacity,
whereas it is shown in section \ref{res} that CN is optically thin.)
According to the models of \citet[][]{BS05}, the above conditions are
consistent with molecular gas at A$_{V} \simeq$ 4 -- 5 mag and a range
of densities and radiation fields, $\chi$.  In the range A$_{V} \sim 4
- 6$ the predicted column ratio changes dramatically \citep[][]{BS05},
so the implied A$_{V}$ is not strongly sensitive to the optical depth
of HCN as long as it is $<$10.  Moderate densities ($\le 10^{4}$
cm$^{-3}$) and radiation fields ($\chi < 10^{3}$), where $\chi$ is the
normalized Draine solar neighborhood radiation field at 1000 \AA\ are
slightly favored since as density and $\chi$ increase, peak CN
abundances drop below $10^{-8}$ as ionization is pushed deeper into
the cloud.  Molecular gas at A$_{V} \simeq$ 4 -- 5 corresponds roughly
to the transition between PDR and dark cloud conditions.  So the
starburst ISM traced by CN and HCN in \gal\ appears to be influenced
by PDRs, but not strongly.  Furthermore, the fact that the
HCN(1--0)/HNC(1--0) ratio $>1$ over the CNZ favors PDR conditions.
Localized chemical effects associated with X-ray dominated regions
such as from a buried AGN are not pronounced \citep[e.g.,][]{MSI07}.

To explain the total observed CN column density when $\rm n_{H_{2}}\le
10^{4}$ cm$^{-3}$ and $\chi \lesssim 10^{3}$ with A$_{V} \sim$ 4 -- 5
gas requires $\sim$45 PDR clouds along the line-of-sight.  Using the
same observed ratio \citet[][]{BS05} and \citet[][]{Fuente+05}
estimate somewhat higher $\chi$'s for the giant PDR in M 82, but fewer
(10 - 20) numbers of PDR clumps.  The number of clumps estimated here
is 2 - 4 times larger than for M 82, the same factor by which the SFR
is larger.

However it is hard to explain why HCN(1--0)/CN(1--0) decreases toward
the outer CNZ in the context of PDR gas.  This implies that the
abundance of the PDR tracer CN relative to HCN increases with distance
from the starburst, contrary to expectation.  One possible explanation
is that the ionization remains high over the entire CNZ, via the
weaker extended star formation component or elevated cosmic ray
ionization rate and the gas away from the starburst being on average
somewhat more diffuse. This model of more diffuse gas in the outer
CNZ, away from the starburst, is also consistent with the lower
HCN(1--0)/CO(1--0) ratio there, since diffuse clouds have lower total
dense gas fractions.
 
\subsection{Gas-Grain Chemistry Across \gal \label{chem}} 

CH$_{3}$OH and HNCO have no efficient gas-phase pathway but are easily
formed on grain mantles.  Large gas-phase abundances of these species
require mantle disruption, either by evaporation or shocks.
CH$_{3}$OH is observed to peak just north of the starburst.  Here
CH$_{3}$OH abundances reach $1.8\times 10^{-8}$ (Table
\ref{MolT}). HNCO is tentatively detected toward the CNZ, with an
abundance of $\sim2.3\times 10^{-9}$.  Both are highly elevated
relative to abundances in quiescent molecular clouds in the Galaxy
\citep[][]{KDBWA97,MB02} and comparable to the abundances observed on
GMC scales toward the strongest shock regions in nearby galaxies
\citep[e.g.,][]{MT05,MT12}.

Since the SFR is large toward the starburst, we investigate whether
the nuclear CH$_{3}$OH intensity can be explained by large collections
of compact sources such as proto-stellar outflows or hot cores
\citep[e.g.,][]{UGMFN06}. We estimate the CH$_{3}$OH intensity
produced from molecular outflows by taking the size and CH$_{3}$OH
brightness of the proto-typical molecular outflow, L1157
\citep[][]{BP97}.  At the distance of \gal, we estimate a single such
outflow will contribute an intensity, $I(CH_{3}OH)_{out} \sim
1.3\times 10^{-6}$ K km s$^{-1}$.  To reproduce the observed intensity
(Figure \ref{IntI}), we require $3.2\times 10^{7}$ such outflows, or
290 outflows pc$^{-2}$.  The surface density of outflows predicted
from a Salpeter IMF (M$_{u}=100$ M$_{\odot}$; M$_{l}=0.1$
M$_{\odot}$), a lifetime of 10$^{-4}$ yr and a SFR of 12 M$_{\odot}$
yr$^{-1}$ \citep[][]{MTBGTV10}, is 3.1 outflows pc$^{-2}$, two orders
of magnitude too low to explain the observed methanol emission. The
situation is even more untenable for HNCO \citep[][]{RTGB10}.  Nor can
hot molecular cores, which are roughly as common as outflows but an
order of magnitude smaller in angular size, match the observed
abundances.

Therefore we conclude that only (large-scale) shocks are capable of
explaining the observed abundances and that CH$_{3}$OH provides a
mapping of those shocks across the galaxy.  This agrees with what is
seen in local starburst galaxies \citep[][]{MT05,MT12}.  Clearly the
CNZ is a site of strong dynamical shocks.  That the CH$_{3}$OH peaks
off of the starburst could reflect shocks associated with the
intersection of the inner molecular arms with the CNZ, as is observed
in nearby bars \citep[][]{MT05,MT12}.  It is also possible that the
weakness of CH$_{3}$OH at the starburst is due to depopulation of the
low lying energy states or to direct photo-dissociation.

Toward the end of both bar arms we tentatively ($\sim 4 \sigma$ in the
north and $\sim 3 \sigma$ in the south) detect CH$_{3}$OH emission.
CH$_{3}$OH abundances toward these locations are comparable to those
seen toward the CNZ.  CH$_{3}$OH is similarly enhanced at the end of
nuclear bar of the nearby starburst Maffei 2, suggesting that the
structure of the (large-scale) bar here resembles the {\it nuclear}
bar in Maffei 2 \citep[][]{MT12}. Shocks have been proposed as a
deterrent to star formation along the arms, allowing gas to drift into
the nuclear starburst region \citep[][]{MTBGTV10}.  CH$_{3}$OH
emission is not detected elsewhere in the inner bar arms, but limits
on its abundance ($< 1.8\times 10^{-8}$) are inconclusive.  So the
observations do not rule out the existence of such strong shocks along
the inner arms.

The tentative detection of HNCO toward the nuclear disk permits a
limit to be placed on the
HNCO($5_{05}-4_{04}$)/CH$_{3}$OH($2_{k}-1_{k}$) line ratio.  As
discussed in \citet[][]{MT12}, the HNCO/CH$_{3}$OH line ratio is
sensitive to the physical conditions present in the shocked gas.  This
ratio is sensitive both to photodissociation and gas density.  HNCO is
both more rapidly photodissociated \citep[e.g.,][]{MMV09} and the
observed transition has a slightly higher critical density than
CH$_{3}$OH.  toward the nuclear disk HNCO/CH$_{3}$OH $\sim$ 0.25,
while toward the northern bar arm, HNCO/CH$_{3}$OH $<$ 0.20.  These
ratios are small compared to those observed on small scales in the
moderate star formation rate nuclear bars, IC 342 and Maffei 2
\citep[0.9 - 1.1;][]{MT05,MT12}, but comparable to the limits in M 82
\citep[][]{MMM06,AMMMHOA11}.  Hence the low HNCO/CH$_{3}$OH ratio
suggests that either densities are on average lower or shocked gas is
more strongly penetrated by UV radiation in the \gal\ CNZ than in the
nuclear bars of normal spiral galaxies.  The brightness of the dense
gas tracers toward the CNZ argues against the former explanation and
the strong starburst favors the latter.  However the latter is in some
tension with the modest implied PDR penetration ($\S$ \ref{pdr}).
Away from the starburst, at the end of the bar, lower gas density may
be the reason for the low HNCO/CH$_{3}$OH intensity ratio.

\subsection{\gal\ Nuclear Physical Conditions Compared to Nearby Starbursts \label{compare}} 

Extensive recent work has focused on the dense gas properties of
nearby, strong starbursts \cite[e.g.,][]{HBM91, NJHTM92, SDR92,
  KMVOSOIK01, APHC02, GS04, GGPC06, PAG07, INTOK07, KNGMCGE08,
  BNSJWSVMW08, BHLBW08, GGPFU08, JNMSBKV09, BAMVM09,
  JWG11,GUAGPCPA12}.  Here we place the dense gas properties of
\gal\ in the context of these studies.

The CNZ intensity ratios, CO/HCN $\sim$ 7, HCN/HCO$^{+}$ $\sim$ 1.0,
HCN/HNC $\sim$ 1.3, and HCN/CN $\sim$ 2.0 fall well within the range
found for other luminous starburst galaxies.  The CO/HCN ratio lies at
the low end of the range, and the HCN/CN ratio lies at the high end of
the range, while HCN/HCO$^{+}$ and HCN/HNC are more typical
\citep[e.g.,][]{BHLBW08}.  The limit on the HC$_{3}$N/HCN line ratio
of 0.086 is very similar to the ratio found for NGC 253
\citep[e.g.,][]{LACPMM11,BHLBW08}.  So either \gal\ does not have a
strongly enhanced very high density gas component or elevated
excitation moving population up out of these intermediate J level
HC$_{3}$N transitions plausibly explains the observed faintness of
this 3mm line \citep[][]{MHS90,MTS11,MT12}.

The dense gas line ratios observed toward \gal's starburst are similar
to those observed with single-dish telescopes toward the starburst in
NGC 253, except that the CO/HCN line ratio in \gal\ is $\sim$3 times
lower than NGC 253's single-dish value and the HCN/CN is a factor of
four larger \citep[e.g.,][]{BHLBW08}.  However the CO/HCN line ratio
toward NGC 253 measured at the $\sim$3\arcsec\ approaches that of
\gal\ \citep[][]{KWWBRM07}.  Therefore the higher single-dish CO/HCN
ratio in NGC 253 is likely an artifact of more diffuse gas being
included in the single-dish beam compared to \gal.  The comparatively
faint CN indicates that the starburst ISM in \gal\ has not been
penetrated by PDRs to quite the degree of NGC 253.

The higher HCN/CN intensity and the higher dense gas fraction suggests
that the starburst in \gal\ is at an early stage of evolution.
\citet[][]{BLS10} generate simple decaying starburst models to track
the evolution of the dense component in a starburst.  They use the
model to simultaneously predict the CO/HCN intensity ratio and
L$_{IR}$.  \gal's CO/HCN and L$_{IR}$ place its burst at the earliest
phases ($\tau_{age} \le 5\times 10^{6}$ years.) This interpretation is
also consistent with the CO isotopologue data.  \citet[][]{MTBGTV10}
have concluded that the starburst is not sustainable, a true burst,
based on the fact that the nuclear gas consumption time is much
shorter than the gas inflow rate along the bar, and that it must be a
young starburst since molecular mass constitutes $\sim 25-30\%$ of the
dynamical mass in the CNZ (Table \ref{GalT}) .

The morphological similarity between NGC 253 and \gal\ is strong.
Both are strongly barred spirals with starbursts triggered by
bar-driven inflow.  The nuclear starbursts in both \gal\ and NGC~253
are close cousins, with \gal's burst being somewhat more intense, at
an earlier phase and residing in a possibly less processed disk.

\section{Conclusions \label{conc}}

We have imaged emission from lines of dense gas tracers at 3mm with
OVRO and CARMA toward the nearby LIRG, IRAS 0496+2923. The images have
3\arcsec\ to 4.5\arcsec\ spatial resolution, and cover the central
arcminute region ($\sim$ 8.4 kpc in diameter). We have combined these
data with archival OVRO CO data to study chemical effects of an
intense nuclear starburst, SFR $\sim 10~\rm M_\odot\, yr^{-1}$, on its
gas.

\begin{enumerate}

\item We confirm that \gal\ is one of the brightest molecular
  line-emitting galaxies in the sky.  We detect significant emission
  in CO(1--0), $^{13}$CO(1--0), C$^{18}$O(1--0) HCN(1--0), HNC(1--0),
  HCO$^+$(1--0), CN(1--0; 3/2--1/2), CN(1--0; 1/2--1/2), and
  CH$_{3}$OH. Gas properties in \gal\ are similar to those in the
  starburst galaxy NGC~253, although \gal\ is more gas-rich.

\item Emission from the dense gas tracers, HCO$^{+}$, HNC, and CN, is
  primarily confined to the inner $R\sim$ 500 pc (5\arcsec)
  circum-nuclear zone (CNZ), and requires densities of $n\gtrsim 2-10
  \times 10^4~\rm cm^{-3}$ here. It is within this component that the
  starburst is located.  The CO/HCN ratio in \gal\ is $\sim 6-8$ for
  the inner $R<500$~pc region, and $\ge 12$ in most locations within
  the arms. This ratio traces the fraction of molecular gas that is
  dense, indicating nearly a factor of two higher dense-gas fraction
  in the nucleus as compared to the arms (though two locations in the
  arms have CO/HCN ratios comparable to the nucleus).

\item The three CO isotopic lines, which trace less dense gas, are
  detected across the central $R\sim 5-6$~kpc ($\sim 40$\arcsec)
  diameter.  \rth, the CO(1--0)/$^{13}$CO(1--0) line ratio, $= 16\pm
  4$ in the CNZ ($R< 500$~pc) of \gal\ and falls to the very low value
  of $1.9\pm0.4$ by the outer end of the bar. The higher-than-Galactic
  \rth\ in the CNZ is consistent with observations of other actively
  star-forming galaxies.  $^{12}$CO(1--0)/C$^{18}$O(1--0), \ret\,,
  follows the same pattern with \ret~$=45\pm10$ in the nuclear
  starburst--CNZ and falling to $3.3\pm0.7$ toward the outer
  disk. This trend is likely due to either lower opacities in CO due
  to kinematics or to higher excitation temperature in the CNZ, which
  removes molecules from the lowest CO rotational levels.

\item The intensity ratio $^{13}$CO(1--0)/C$^{18}$O(1--0) or \rthet
  ~is low, \rthet~$=3.2\pm0.8$, in the CNZ, and even lower in the
  outer bar, with \rthet $\lesssim 2$.  Interpretation of this
  unprecedentedly low ratio in terms of very high CO opacities
  conflicts with the high observed CO(1--0)/$^{13}$CO(1--0).  We can
  obtain barely consistent solutions for the CO isotopic ratios if we
  adopt [CO]/[$^{13}$CO] $\sim$ 60 toward the nucleus of \gal,
  increasing to 110 at the bar ends, and [CO]/[C$^{18}$O]$\sim$200 --
  225.  These values may indicate a relative lack of long term stellar
  processing in \gal\ as compared to the Galaxy.  For the implied
  isotopic abundance ratios, optically thin C$^{18}$O(1--0) emission
  predicts a nuclear conversion factor $\sim$0.5 times that of the
  Galactic disk.  The implied conversion factor in the bar arms are
  approximately equal (or slightly larger) than in the Galactic disk,
  consistent with that found in \citet[][]{MTBGTV10}.
 
\item The observed global value of $L_{IR}/L_{HCN}\simeq1300$
  L$_{\odot}$ (K km s$^{-1}$ pc$^{2}$) obtained for \gal\ is nearly a
  factor of 1.5 times the standard global value for the IR to HCN
  luminosity ratio obtained by \citet{GS04}.  Excluding the non-CNZ IR
  luminosity, we obtain a value for the CNZ alone of $L_{IR}/L_{HCN}
  \simeq 660$ L$_{\odot}$ (K km s$^{-1}$ pc$^{2}$) for \gal.  From
  this we estimate a star formation rate relation $\rm
  SFR(M_\odot\,yr^{-1})=1.4\pm0.3 \times 10^{-7}~L_{HCN}$, similar
  given the large systematic uncertainties, to that obtained by Gao \&
  Solomon.

From LTE analysis we find that the CN abundance is $\sim$1.6 times the
abundance of HCN, if HCN is optically thin.  If HCN is optically thick
then HCN can still be more abundant than CN. Based on the models of
\citet[][]{BS05} the HCN/CN column density ratio indicates that the
emission is coming from clouds at moderate depths of $A_v\sim 4$.
This corresponds to the transition between PDRs and dark clouds.  So
the CNZ is moderately influenced by PDRs.  The HCN/CN ratio decreases
radially from the starburst in the CNZ.  This may suggest that
PDR-influenced gas extends well beyond the compact starburst coupled
with a drop in characteristic A$_{v}$.

\item Bright emission from the CH$_3$OH molecule indicates that grain
  chemistry is important on large scales in \gal. We are unable to
  reproduce the emission from models of collections of compact
  sources, and conclude that only a widespread mechanism such as
  shocks along spiral arms can explain the observed brightnesses of
  these species.  A tentative detection of CH$_3$OH at a
  galactocentric radius of $>$2 kpc is presented.  The CH$_{3}$OH
  abundance in this region are comparable to those found in the CNZ.

\item All lines of evidence, including the unusual CO isotopologue
  ratios and CO/HCN vs. $L_{IR}$, indicate that the starburst in
  \gal\ is very young, which is consistent with previous suggestions
  based on bar-fueling of the starburst \citep{MTBGTV10}.

\end{enumerate}

\acknowledgements 
 
DSM acknowledges support from the NSF under grant AST-1009620.  We
thank the anonymous referee for a thorough and helpful review. Support
for CARMA construction was derived from the states of California,
Illinois, and Maryland, the James S. McDonnell Foundation, the Gordon
and Betty Moore Foundation, the Kenneth T. and Eileen L. Norris
Foundation, the University of Chicago, the Associates of the
California Institute of Technology, and the National Science
Foundation.  Ongoing CARMA development and operations are supported by
the National Science Foundation under a cooperative agreement, and by
the CARMA partner universities.

{\it Facilities:} \facility{CMA}, \facility{OVRO}

\end{document}